\documentstyle[12pt,epsf]{article}

\newcommand{\xbj}{x_{B}}
\newcommand{\dt}{\Delta_T}

\newcommand{\M}[4]{M^{#1, #2}_{#3, #4}}     
\newcommand{\tvc}{{\cal T}_{\mathit{VCS}}}     
\newcommand{\tbh}{{\cal T}_{\mathit{BH}}}      
\newcommand{\eqpt}{\hspace{6pt}.\hspace{6pt}}  
\newcommand{\eqcm}{\hspace{6pt},\hspace{6pt}}

\newcommand{\gstar}{\gamma^\ast}

\newcommand{\weight}[1]{\langle\langle \, #1 \, \rangle\rangle}

\begin{document}

\title{Deeply Virtual Compton Scattering:\\
How to Test Handbag Dominance?}
\author{T. Gousset$^{a,b}$, M. Diehl$^{c,d}$,
B. Pire$^c$ and J.P. Ralston$^e$\\
{\small\it $a)$  NIKHEF, P. O. Box 41882, 1009 DB Amsterdam, The 
Netherlands.}\\
{\small\it $b)$ SUBATECH, B. P. 20722, 44307 Nantes CEDEX 3, France.}\\
{\small\it $c)$ CPhT, Ecole Polytechnique, 91128 Palaiseau, France.}\\
{\small\it $d)$ DAPNIA/SPhN, C. E. Saclay, 91191 Gif sur Yvette, France.}\\
{\small\it $e)$ Dept.\ of Physics and Astronomy, University
of Kansas, Lawrence, KS~66045, USA.}}
\date{}
\maketitle

\begin{abstract}
We propose detailed tests of the handbag approximation in exclusive
deeply virtual Compton scattering. Those tests make no use of any
prejudice about parton correlations in the proton which are basically
unknown objects and beyond the scope of perturbative QCD. Since
important information on the proton substructure can be gained in the
regime of light cone dominance we consider that such a class of tests
is of special relevance.
\end{abstract}

\section{Why deeply virtual Compton scattering\dots}

There has recently~\cite{Ji,Rad1,CHR} been considerable interest in
exclusive virtual Compton scattering (VCS) and the proposal that
off-diagonal matrix elements of quark operators in proton states
might be measurable there. The idea is that the process
\begin{equation}
\gstar(q)+p(p)\to\gamma(q')+p(p')
\end{equation}
proceeds via the short-distance Compton scattering on a single quark
(cf.\ Fig.~1) in the Bjorken limit of large $Q^2=-q^2$ at fixed
$\xbj=Q^2/(2p\cdot q)$, and for a limited range of the momentum
transfer $t = (p-p')^2$, or in other words of the transverse momentum
$\dt$ of the scattered proton (in the c.m.\ of the collision), which
should be of the order of a hadronic scale. (Through a hard quark loop
the two photons can also couple to two gluons.) Following the old
terminology of deep inelastic scattering those diagrams where two
quark/gluon fields are connected to the proton states are referred to
as {\em handbag} diagrams. The short-distance processes can be
computed in perturbation theory and the VCS amplitude is a sum of
convolutions of those perturbative amplitudes with the Fourier
transforms of off-diagonal matrix elements of quark or gluon fields
such as (in the $A^+=0$ gauge)
\begin{equation}
\langle p'|\bar{\psi}(0)\gamma^{+}\psi(z)|p\rangle \eqcm\ 
\langle p'|G^{+i}(0)\,G^{+}{}_{i}(z)|p\rangle  \eqcm
\end{equation}
quantities that have been introduced in the literature some time ago
(for references cf.~e.g~\cite{us}).  We note that handbag dynamics
singles out both specific Dirac/Lorentz components of the quark/gluon
fields and light-like separations between them. At large $Q^2$ other
contributions including correlations of three or more fields are power
suppressed.

\begin{figure}[ht]
$$\epsfysize 32mm\epsfbox{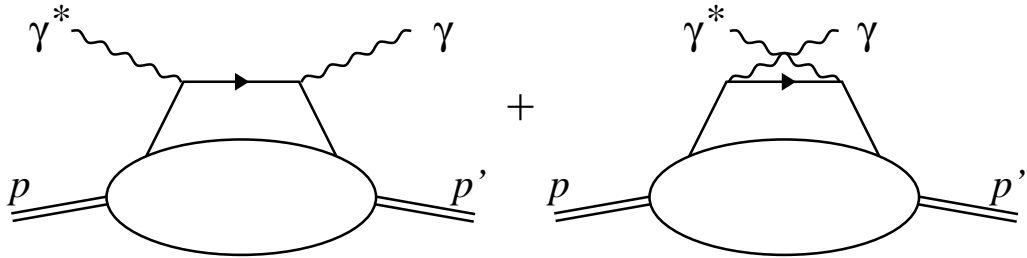}$$
\caption{The handbag Born diagrams for virtual Compton scattering.}
\end{figure}

{}From this line of reasoning it is clear that the process shares
similarities with inclusive deep inelastic scattering, whose cross
section is directly proportional to the imaginary part of the forward
Compton amplitude. There one also selects in the Bjorken limit
products of two quark or two gluon fields with light-like
separations. The Compton scattering reaction is however richer because
of the difference between the initial and final state protons. The
light-like separation being probed in a different manner may help us
to extend our understanding of proton structure at the quark-gluon
level.

A very interesting feature discovered by Ji~\cite{Ji} is that sum
rules for the off-diagonal matrix elements carry information on the
orbital angular momentum of quarks and gluons in the proton. From this
point of view their study naturally extends that of proton spin
structure, which has been an object of intense research in the last
decade.

\section{\dots and how?}

There is now some theoretical evidence for the separation of long and
short distance dynamics in the framework of perturbative QCD. An
important question is then to ask how to test this prediction
experimentally. We have in mind tests which rely on general principles
and avoid theoretical prejudices or model assumptions about the
unknown off-diagonal matrix elements.

To formulate tests of the dominance of the handbag diagrams we make
use of their spin selection rules for the $\gstar p\to\gamma p$
process. For this purpose we consider the perturbative Compton
scattering of a virtual photon on a free quark target in the high
energy limit. In a frame where all particles are fast and in the
region of nearly collinear kinematics we are considering, the process
has the remarkable property of helicity conservation for the
photon. This follows from the fact that a massless quark cannot change
its helicity. (QCD loop corrections to order $\alpha_S$ may violate
the photon helicity conservation rule by two units, but interestingly
not by one~\cite{us}.)  At $\dt = 0$ the proton helicity can then not
be flipped either because of angular momentum conservation. For finite
small $\dt$, however, the handbag {\em does} give proton helicity flip
amplitudes at leading twist, with an overall factor of $\dt R$, where
$R$ is of the order of the proton radius.

The reaction where one can study virtual Compton scattering is the 
electroproduction process
\begin{equation}  \label{ep}
  e(k) + p(p) \to e(k') + p(p') + \gamma(q') \eqpt
\end{equation}
A great opportunity is offered by the interference of VCS with the
Bethe-Heitler (BH) amplitude, where the photon is radiated off the
electron. In addition to the kinematical invariants introduced to
describe VCS we use the standard variable $y=(q\cdot p)/(k\cdot p)$
and the angle $\varphi$ between the leptonic and hadronic scattering
planes in the c.m.\ of the scattered photon and scattered proton. The
$\varphi$-dependence of the $ep$ cross section contains a wealth of
information, as we shall see.

With appropriate phase conventions for the external particles the
contribution of VCS to the amplitude of~(\ref{ep}) for a given
helicity $\lambda$ of the intermediate photon $\gstar(q)$ depends on
$\varphi$ as
\begin{equation}
  \label{VCSamplitude}
  \tvc \propto \exp(- i \lambda \varphi)  \eqpt
\end{equation}
The BH amplitude has in general a rather complicated
$\varphi$-dependence, but at leading order in $1 /Q$ it simplifies to
\begin{equation}
  \label{BHamplitude}
  \tbh \propto \exp(- 2 i \lambda' \varphi)  + O\left( 1 \over Q
  \right)
\end{equation}
for a scattered photon of helicity $\lambda'$. With this one can
analyze the $\varphi$-dependence of the $ep$ cross
section~\cite{us}. It displays a hierarchy in powers of $1/Q$, with $|
\tbh |^2$, BH--VCS interference and $| \tvc |^2$ respectively going
like 1, $1/Q$ and $1/Q^2$ in the handbag model, accompanied with a
$\varphi$-distribution going like $\cos (n \varphi)$ with $n = 0$, 1,
2 and 3. A powerful consequence is that extracting the angular
distribution does not require any binning of data. Using the weighted
cross sections
\begin{equation}
  \label{weighted}
\weight{f(\varphi)} = \frac{Q^4}{\xbj \, y^2} \, \int d\varphi \,
  {d\sigma \over d\varphi \, d t \, d Q^2 \, d\xbj} \cdot f(\varphi)
\end{equation}
with $f(\varphi)= 1, \cos\varphi, \cos 2\varphi, \cos 3\varphi$, one
directly projects out the coefficients of $\cos (n \varphi)$ in $|\tvc
+ \tbh|^2$ and can formulate a number of tests for the predictions of
the handbag approximation. We restrict ourselves here to the case
where $y$ is not too small; then the BH--VCS interference dominates
over the squared VCS amplitude.

We can test the scaling properties of the handbag. Let us define
helicity amplitudes $\M{\lambda}{\lambda'}{h}{h'}$ for $\gstar p \to
\gamma p$, where $\lambda$ ($\lambda'$) is the helicity of the initial
(final) state photon and $h$ ($h'$) that of the initial (final) state
proton. To order $1 /Q$ the $\cos\varphi$-term in the BH--VCS
interference is multiplied by the photon helicity conserving
amplitudes $\M{1}{1}{h}{h'}$ which the handbag approximation predicts
to be constant in $Q$. There is also a $\cos\varphi$-term in the
square of the BH amplitude, which has the same global power $1 /Q$ and
thus must be subtracted if we want to investigate VCS. Note that the
BH process including its QED radiative corrections can be calculated,
and that the elastic proton form factors are well parameterized in the
region of small $t$ where they are needed. We then have as a test for
the scaling properties in the handbag that at fixed $\dt$, $\xbj$ and
$y$
\begin{equation}
  \label{cos}
  \weight{\cos\varphi} -
  \left.\weight{\cos\varphi}\right._{\mathrm{BH~only}} \sim 1 /Q  \eqcm
\end{equation}
where $\smash{ \left.\weight{\cos\varphi}\right._{\mathrm{BH~only}} }$
denotes the contribution of the squared BH amplitude. Of course this
scaling behavior is to be understood as up to logarithms due to QCD
radiative corrections in the handbag.

To test photon helicity conservation we can use $\weight{\cos
2\varphi}$ and $\weight{\cos 3\varphi}$. To order $1 /Q$ they receive
contributions from the BH--VCS interference proportional to
$\M{0}{1}{h}{h'}$ and $\M{-1}{1}{h}{h'}$, respectively, which are zero
in the handbag approximation and thus should be power
suppressed. $|\tbh |^2$ does not contain any $\cos 2\varphi$ or $\cos
3\varphi$ up to order $1 /Q$ so that, {\sl without needing to subtract
this contribution}, we have as a test for the handbag that
\begin{equation}
  \label{costwothree}
  \weight{\cos 2\varphi} \eqcm \weight{\cos 3\varphi} \sim 1 / Q^n
  \eqcm n \ge 2  \eqpt
\end{equation}
{}From (\ref{cos}) and (\ref{costwothree}) one has of course that
$\weight{\cos 2\varphi}$ and $\weight{\cos 3\varphi}$ should be small
compared with $\weight{\cos\varphi}$, a test that can even be done
without much lever arm in $Q^2$.

For lack of space we shall not discuss here the kinematic regime where
$y$ is close to zero and where the VCS contribution to the amplitude
dominates over the contribution from BH. Additional and complementary
information on VCS can be gained using lepton beams with longitudinal
polarization, while still averaging over the proton spin; the
interested reader is referred to Ref.~\cite{us}.

\section*{Acknowledgments}
This work is partially funded through the European TMR Contract
No.~FMRX-CT96-0008 and through DOE grant No.~85ER40214. T. G. 
was carrying out his work under the European Contract No.~ERBFMBICT950411.
CPhT is Unit\'e propre 14 du Centre National de la Recherche Scientifique.


\begin{thebibliography}{99}
\bibitem{Ji} X. Ji, Phys.\ Rev.\ Lett.\ {\bf 78}  610 (1997); Phys.\
  Rev.\ {\bf D55} 7114 (1997).
\bibitem{Rad1} A.V. Radyushkin, Phys.\ Lett.\ {\bf B380}  417 (1996);
Phys.\ Rev.\ {\bf D56} 5524 (1997).
\bibitem{CHR} C. Hyde-Wright, Nucl. Phys. {\bf A622} 268c (1997).
\bibitem{us} M. Diehl, T. Gousset, B. Pire and J.P. Ralston, Phys.\
Lett.\ {\bf B411} 193 (1997).
\end{thebibliography}
\end{document}